Polarization dependence of photo-mechanical behavior of monodomain liquid crystal polymeric materials

Chen Xuan[1], Changwei Xu, Yongzhong Huo[2*]

Department of Mechanics and Engineering Science, Fudan University, Shanghai 200433, China

**Abstract**

Polarization dependence of opto-mechanical behavior of monodomain photochromic glassy liquid crystal (LC) polymers under polarized ultraviolet light (PUV) is studied. *Trans-cis* photo-isomerization is generally known to be most intense at "parallel illumination" (polarization parallel to LC director), as light-medium interactions are active when polarization aligns with trainsition dipole moment. We show that at parallel illumination though *cis* isomers are converted from *trans* the most near surface, they can be the least below certain light propagation depth. Membrane force, an average effect of *trans-cis* conversion over propagation depths, shows a monotonic polarization dependence, *i.e.* maximum at parallel illumination, which agrees well with experiment [1]. However, under strong illumination, *cis* fraction/photo-contraction distribution through depths shows deep penetration, switching over the polarization dependence in photo-moment, which is related to photo-contraction gradient ---- photo-moment can be maximum at "perpendicular illumination" (polarization perpendicular to director) under strong light. We give both intuitive explanation and analytical demonstration in thin strip limit for the switchover.

1. **Introduction**

Photochromic liquid crystal networks doped with moieties such as azo-dyes contract along their director $n$ when their orientational order is reduced by photo-isomerization [2]-[7]. Such photo-contraction can be very big for soft elastomers but small for hard glass. We focus on the latter, since their directors are nearly frozen, free of further complex nonlinear mechanical response. Two different photo-isomerization processes: *trans-cis* and *trans-cis-trans* are widely studied. The difference between the two is that in *trans-cis-trans* process, *cis* recovering to *trans* is optical rather than thermal as in *trans-cis*. *Trans-cis* is considered in this paper because it is more basic yet still lack of deep understanding when polarization comes into play. More and more studies show high interest on PUV in inducing and controlling photo-mechanical response [6][8][14]. In particular, photo-bending direction in polydomain films can be controlled by polarization [6]; photo-contraction or membrane force is maximum when polarization aligns with director [1]. It is well known that photo-isomerization happens particularly actively at parallel illumination, as optic behavior is intense when polarization is parallel to molecule transition dipole moment [15][16]. Modeling of non-polarization dependent photo-mechanics in cantilevers, strips, plates, films can be seen in, say [17]-[22]. However, detailed *trans-cis*-induced photo-response in glassy LCs in terms of photo-strain distribution through light propagation depths, patterns of photo-bending curvature are still unclear, especially regarding their dependence on polarization direction and illumination intensity. PUV triggers reorientation of nematic directors in polydomain LC elastomers to reduce elastic energy [23], where uniform photo-contraction through depths is assumed. This invariably leads us to wonder a basic and seemingly simple question: the exact photo-strain distribution through depths and bending curvature patterns in glassy LCs, with immobile directors. Yet we will show that these are not as trivial as they may seem.

---

[1] Present address: Cavendish Laboratory, University of Cambridge, 19 JJ Thomson Avenue, Cambridge CB3 0HE, United Kingdom. Email: cx226@cam.ac.uk
[2] Corresponding author: yzhuo@fudan.edu.cn

Light-induced strain is expected to be larger at parallel illumination, as this is when light acts most effectively on isomerization. However, this only holds true near illumination surface. Light is better absorbed at parallel illumination ---- stronger light decay, that is parallel illumination bleaches (so named because the converted dye is not an effective absorber of the original color of light when in the *cis* state) *trans* most effectively near surface but makes *cis* fraction or photo-contraction decay with depth most rapidly (in Beer limit, parallel illumination reduces Beer penetration length). In addition for a particular sample, whether or not *trans* bleaching front lie in (shallow light penetration) or out (deep light penetration) of the sample determines photo-contraction distribution and its photo-moment. A possible intuition would be that curvature or photo-moment has the same polarization dependence as membrane force does in [1]. However as shown in [18], their behavior can be rather different, particularly at large light intensities, with strong nonlinearities involved. As photo-moment depends on details of photo-contraction distribution, its polarization dependence could be more involved on top of its own nonlinearities.

In this paper we adopt the polarization dependent *trans-cis* isomerization theory proposed in [16] to study photo-contraction depth profile and bending moment in glassy LC polymers. The main motivations and differences from literature are ---- (i) Polarization dependent *trans-cis* isomerization in glassy LCs photo-response is not fully clear and so requires thorough investigation. (ii) *Cis* fraction and photo-strain depth profiles are lack of research regarding their dependence on polarization and illumination intensity. (iii) Photo-moment or curvature would show unusual phenomena originated from the photo-strain depth profile ---- To eliminate complexities as much as possible we concentrate on propagation depths much thinner than depolarization length, in a limit where polarization direction does not vary with depth in the LC anisotropic medium. This assumption would be deficient in detail for large depths and thick strips in that the incident light should have been decomposed into ordinary and extraordinary components, with different refractive indices in an anisotropic medium. However, we point out that our limit is accurate in considering parallel illumination and perpendicular illumination. Whether it be out of optical analysis or geometrical symmetry, polarization direction will not vary with depth in both illuminations. All of the results for parallel and perpendicular illuminations in this paper, which are accurate, insure the validity of various switchovers discovered, even though the quantitative results for polarization at any random angles to director are approximate. Finally, we suggest experiments to verify to what extent our limit properly explains.

## 2. Polarization dependence of photo-isomerization at different propagation depths

We consider linearly polarized ultraviolet light (with polarization direction $e$) illuminated normal onto a monodomain glassy LC medium (with director $n$), causing *trans-cis* reaction alongside with *cis-trans* thermal back reaction. The conversion from *trans* to *cis* increases proportionally with remaining *trans* fraction $n_t$ and light intensity $I$ entered, but decreases proportionally with existent *cis* fraction $n_c$. Meanwhile, light intensity $I$ decays when propagating through the absorbing medium, satisfying a modified Lambert-Beer's Law, that is $I$ decays proportionally with $n_t$ and $I$ itself. To ignore possible depolarization of light due to the anisotropy of LCs, we assume the same ansatz as in [16][23][24] that penetration depths we consider in this paper is small enough compared with depolarization length such that the transmitted light always keeps its polarization direction the same as that of the incident light. The $n_c$ and $I$ are coupled in the following partial differential equations [16][23][24] in dynamics.

Though dynamics of *trans-cis* isomerization could be interesting both experimentally and theoretically, we study photo-stationary state and its effects on bending mechanics in this paper for theoretical convenience. This approach is not lack of experimental merit, because photo-stationary state can always be expected as one waits for long enough, whereas instantaneous observation might take more efforts. So when the isomerization process progresses long enough one gets

$$n_c(z) = \frac{\tau_{ct}\eta I(z)}{1+\tau_{ct}\eta I(z)}. \quad (1)$$

where $\tau_{ct} = \exp(\Delta/k_B T)$ is the characteristic time for *cis-trans* thermal back transition. As easily seen from the above equation, $\tau_{ct}(T)$ has similar contributions as $I$ does to $n_c$. Even though temperature might be slightly influenced by photo-isomerization, we assume that the temperature is held constant (thermal equilibrium quickly achieved) by the environment temperature at the reference temperature, $T=T_0$, for simplicity throughout the rest of the paper. Albeit LC order parameter for glassy nematics could be changed a small amount by photo-isomerization, we found that this does not give a significant quantitative difference for the *cis* fraction. Therefore the order parameter here can be roughly treated as a constant at a given temperature (In the figures afterwards, the order parameter $Q$ is set as 0.5 unless otherwise stated). Thus, there is an explicit solution expressed by Lambert-W function (ProductLog function), where $W(x)$ is defined as $x=W(x)e^{W(x)}$[20]

$$I = \frac{1}{\tau_{ct}\eta} W[\tau_{ct}\eta I_0 e^{(\tau_{ct}I_0 - \gamma_t z)\eta}], \text{ with } \eta = \eta(\psi; Q), \quad (2)$$

$$n_c = \frac{W[\tau_{ct}\eta I_0 e^{(\tau_{ct}I_0 - \gamma_t z)\eta}]}{1+W[\tau_{ct}\eta I_0 e^{(\tau_{ct}I_0 - \gamma_t z)\eta}]}, \text{ with } \eta = \eta(\psi; Q), \quad (3)$$

where $I_0$ is the light intensity incident onto illumination surface, $i_0 = \tau_{ct}(T_0)\eta_0 I_0$ a dimensionless light intensity, $\gamma_t$ a positive material constant. One can define a characteristic penetration or absorption length $d = (\gamma_t \eta_0)^{-1}$. $\eta$ is an absorption coefficient to describe polarization dependence [16][23][24]

$$\eta(\psi; Q) = \eta_0[1+2QP_2(\cos\psi)] = \eta_0[1+Q(3\cos^2\psi - 1)] \geq 0, \text{ with } \cos\psi = \mathbf{n}\cdot\mathbf{e}, \quad (4)$$

where $\eta_0$ is a positive material constant, $\psi$ (which we will call polarization angle for the rest of the paper) the relative angle between the director *n* and the polarization *e*, and $Q$ order parameter.

Obviously $\eta$ is periodic in $\psi$ with period $\pi$. $\eta$ has maximum $\eta_M = \eta_0(1+2Q)$ at $\psi = 0$, and minimum $\eta_m = \eta_0(1-Q)$ at $\psi = \pm\pi/2$. $P_2(\cos\psi) = (3\cos^2\psi - 1)/2$ is the second Legendre polynomial and decreases monotonically in $\psi \in [0, \pi/2]$. So $\eta$ is monotonically decreasing in $\psi \in (0, \pi/2)$ and has symmetry $\eta(\psi) = \eta(-\psi)$. $\psi$ serves as an input parameter as one controls the incident illumination, but is assumed to be constant through the LC medium in this paper. For more accurate analysis $\psi$ should be considered as a function of propagation depth. For our aforementioned reason, we ignore this optic complexity. Eq. (4) is based on the mechanism that photons are most absorbed when polarization aligns with molecule transition dipole moment. This expression of $\hat{\eta}$ states: (i) Photo-isomerization happens most intensely when $e \parallel n$ and least when $\mathbf{e} \perp \mathbf{n}$. (ii) $\hat{\eta}$ is linear in $Q$, that is order parameter magnifies polarization dependence unless $Q=0$, which will be confirmed in detail in Section 3.2. Parallel illumination maximizes both the growth rate of *cis* fraction and the reduction rate of light intensity with depth. It is noteworthy that when $\mathbf{e} \perp \mathbf{n}$, $\eta_m = \eta_0(1-Q) > 0$ unless $Q=1$, that is photo-isomerization still takes place under perpendicular illumination. We will confirm in Section 3.2 that membrane force and bending moment are non-zero, though small, under perpendicular illumination.

It is not hard to derive from these equations some of the function properties, say: (i) $I$ is monotonically reduced with $z/d$ and $\hat{\eta}$. (ii) $n_c$ does decay with $z/d$ but the dependence of $n_c$ upon $\hat{\eta}$ is not monotonic ---- $n_c$ increases with $\hat{\eta}$ at small $z/d$ and decreases with $\hat{\eta}$ at big $z/d$.

We illustrate the above polarization dependence of $I$ and $n_c$ obtained by (2) and (3). For $I$, it is as clear and simple as one might expect (Fig. 1) ---- $I$ decays with $z/d$ most intensely when $\psi = 0$, *i.e.* $\mathbf{e} \parallel \mathbf{n}$. $n_c$ in Fig. 2a is more thought-provoking. When incident light is relatively weak, say $i_0=0.5$, $n_c$ is close to Beer limit, that is $n_c$ decays almost exponentially with depth $z/d$. Parallel illumination ($\psi = 0$) gives a biggest $n_c$ at the surface, but also reduces it with depth fastest ---- Beer length shortened as stated above. Under certain depths $z/d$, $n_c$ becomes instead minimum when $\psi = 0$, so $n_c$ curves under different $\psi$ intersect as shown. Fig. 2b shows a different view of $n_c$, as a function of $\psi$. There are two critical depths $z^c_{1,2}$ such that: for $z < z^c_1$, $n_c$ is maximum at $\psi = 0$ and decreases monotonically to minimum at $\psi = \pi/2$; for $z^c_1 < z < z^c_2$, $n_c$ is non-monotonic and is maximum somewhere in $0 < \psi < \pi/2$; for $z > z^c_2$, $n_c$ is minimum at $\psi = 0$ and increases to maximum at $\psi = \pi/2$. The two critical depths can be obtained by setting $\partial^2 n_c / \partial \psi^2 (\psi = 0) = 0$ and $\partial^2 n_c / \partial \psi^2 (\psi = \pi/2) = 0$, respectively, and one gets

$$\frac{z_1^c}{d} = i_0 + \frac{1}{1+2Q}$$
$$\frac{z_2^c}{d} = i_0 + \frac{1}{1-Q} \quad . \tag{5}$$

The above switchover of polarization dependence of $n_c$ is resulted from the interplay between light intensity and polarization. Parallel illumination facilitates both photon absorption and *trans* depletion. Although parallel illumination depletes *trans* the most at the surface, light is most absorbed and reduced, which in turn acts against *trans* depletion. Despite a high *cis* fraction at the surface, $n_c$ will eventually drop once light needed for triggering isomerization is extinguished. It is important to point out that this feature always exists independent of illumination intensities.

Apart from the switchover of polarization dependence, $n_c$ for $\psi = 0$ at weak illumination in particular shows an apparent bleaching front (Fig. 2a), above which *trans* is highly depleted and below which the opposite. For big $\psi$, namely *e* far from parallel to *n*, the decay slope of $n_c$ is subdued (Beer length elongated) and so might no longer call it a bleaching front. When incident light is strong, say $i_0=5$, $n_c$ deviates far apart from Beer limit. $n_c$ is big from $z/d=0$ to rather big $z/d$ ---- *trans* depletion is deeply penetrated. Then comes the bleaching front before reaching low $n_c$ layers. When $\psi = 0$, most *trans* is depleted above the bleaching front and least *trans* is depleted below the bleaching front. The main difference brought to the photo-isomerization by weak and strong illumination is that for the former bleaching front starts right from the surface while for the latter it sets in below certain *trans* depletion layers. Alongside with polarization dependence coming into play, all this will consequently lead to a counter-intuitive pattern in bending curvature in Section 3.3.

Photochromic LC glassy networks respond to light by contracting along its director *n*. The photo-contraction can be approximated as a positive material constant $\gamma_\varepsilon$ times the *cis* fraction [14]

$$\varepsilon = \gamma_\varepsilon n_c . \tag{6}$$

Due to the trivial linear relation, all of the above analysis on $n_c$ can be directly applied to photo-contraction $\varepsilon$, that is features such as switchover of polarization dependence and photo-contraction fronts should be fully expected in $\varepsilon$. This photo-contraction is a rather unusual distribution of strain not commonly observed in ordinary materials.

## 3. Polarization dependent photo-contraction and bending of monodomain LC glassy strips

### 3.1 *Membrane force and photo-moment*

Consider a thin LC glassy strip as shown in Fig. 3 with its length *L* much longer than its other two dimensions (thickness *h* in *z* direction and width *w* in *y* direction) ---- the simple beam bending theory [18] applies. We consider director *n* aligns at the long axis (*x* direction) of the strip as in Fig. 3. Light is illuminated from above (-*z* direction), with polarization *e* lying in the *x-y* plane as a controlled variable, an angle $\psi$ to *n*. As incident light triggers *trans-cis* isomerization, the sample contracts along *n* (*x* direction) with an amount of $\varepsilon(z)$. Unlike [18], in this paper we focus on the effect of polarization upon photo-induced membrane force and bending curvature, so we rule out non-uniform illumination along *x* and any mechanical loads.

The total (real) strain along *x* is the addition of photo-strain ($-\varepsilon(z)$) and elastic strain. Assuming small deflection and small strain, the stress along *x* is proportional to the elastic strain satisfying Hooke's law. Researches like [1] examine photo-stress or membrane force, with LC samples illuminated with both ends clamped and strip length held constant. Proved in [19], this kind of boundary condition prohibits bend as well as total strain, leading to a trivial uniaxial tension (membrane force) along the axis of the strip to act against photo-contraction

$$F = \int_0^h E\varepsilon(z)dz,  \quad (7)$$

where *E* is the effective Young's modulus. For glassy LCs, the change of elastic modulus *E* caused by photo-isomerization is too small to be important [18]. So for simplicity, we assume *E* to be a constant and consider a dimensionless average membrane force $f = F/Eh = \bar{\varepsilon} = \int_0^1 \varepsilon(h\xi)d\xi$, which is equal to $\bar{\varepsilon}(h)$, the average photo-contraction over depths.

Because of light decay with *z*, this photo-contraction $\varepsilon$ decays with *z* as well (Fig. 2a) and will cause bend in the *x-z* plane. In the simple beam deflection theory, real (total) strain must be a linear distribution through thickness direction. As photo-contraction is nearly an exponential-type strain distribution (see Fig. 2a) (exactly an exponential distribution in Beer limit) as opposed to the required linear distribution, no real strain is compatible to match photo-strain as it bends, and thus residual stress exists ---- no stress free configuration, though force free, which means stress integral is zero. The means to measure the degree of bend or curvature is effective photo-moment ---- a bending moment needed to balance the photo-bend. Tip displacement is an observable in photo-sensitive LC experiments used as a measure of curvature of nematic cantilevers. From [18], the photo-contraction with a gradient in *z* gives rise to an effective photo-moment

$$M_{eff} = -w\int_0^h E\left[\varepsilon(z) - \bar{\varepsilon}\right]zdz. \quad (8)$$

Similarly, a dimensionless effective photo-moment is defined as $m_{eff} = M_{eff}/Ewh^2 = \int_0^1 \left[\bar{\varepsilon} - \varepsilon(h\xi)\right]\xi d\xi$. This dimensionless photo-moment $m_{eff}$ is proportional to bending curvature, even in a dynamical system but where inertia can be ignored, say in the creeping motion seen by [6].

*3.2 Analytical results under weak light illuminations*

When light is weak, Beer limit applies. Both *I* and $n_c$ are exponential in *z/d*, and $n_c$ is linear in $i_0$. With (6), analytical expressions proportional to $i_0$ can be obtained for *f* and $m_{eff}$ under weak light limit

$$f = i_0\gamma_\varepsilon \frac{d}{h}(1 - e^{-\hat{\eta}h/d}), \quad (9)$$

$$m_{eff} = i_0\gamma_\varepsilon \frac{d^2}{\hat{\eta}h^2}[-1 + \frac{\hat{\eta}h}{2d} + (1 + \frac{\hat{\eta}h}{2d})\exp(-\frac{\hat{\eta}h}{d})]. \quad (10)$$

$\hat{\eta}$ above is Eq. (4). Thus both *f* and $m_{eff}$ are periodic functions of $\psi$ with period $\pi$.

By taking the derivative of (9) with respect to $\psi$, $f$ decreases monotonically within $\psi \in [0, \pi/2]$. According to (4), $f$ has maximum at $\psi = 0$ and minimum at $\psi = \pi/2$ below

$$\begin{cases} f(\psi = 0) = i_0 \gamma_\varepsilon \dfrac{d}{h}[1 - \exp(-\dfrac{h}{d}(1 + 2Q))] \\ f(\psi = \dfrac{\pi}{2}) = i_0 \gamma_\varepsilon \dfrac{d}{h}[1 - \exp(-\dfrac{h}{d}(1 - Q))] \end{cases}. \quad (11)$$

From the above equation, $f(\psi = \pi/2)$ will be zero only for perfectly aligned LCs, *i.e.* $Q=1$, hardly obtainable in experiments. This is why in general ($Q<1$) we observe a tensile membrane force at perpendicular illumination ($\mathbf{e} \perp \mathbf{n}$) for samples with both ends clamped as done *e.g.* in [1].

It is easy to use the explicit expression (9) to fit the measured engineering photo-stress of Fig. 8 in [1]. The engineering stress in [1] is exactly the Young's modulus times (9). The Young's modulus $E$ is given as $3.8 \times 10^4$ Pa in [1]. In (9) there are three parameters: $i_0 \gamma_\varepsilon$, $d/h$ and $Q$. We take the average values for $\psi = 0$, $\psi = \pi/4$ and $\psi = \pi/2$ from the data points in [1] to determine the above three parameters. Fig. 4 shows the original data points in [1] and our fitting curve, which match quite well. The corresponding parameters for the fitting curve are $i_0 \gamma_\varepsilon = 0.14, d/h = 2.82, Q = 0.48$.

The photo-moment (10) is also a decreasing function of $\psi \in [0, \pi/2]$ as can be observed by its following derivative

$$\frac{dm_{eff}}{d\psi} = -3i_0 \gamma_\varepsilon Q \left(\frac{d}{\hat{\eta}h}\right)^2 \exp(-\frac{\hat{\eta}h}{d}) \left[\exp(\frac{\hat{\eta}h}{d}) - 1 - \frac{\hat{\eta}h}{d} - \frac{1}{2}\left(\frac{\hat{\eta}h}{d}\right)^2\right] \sin 2\psi. \quad (12)$$

It is not difficult to show that the minimum photo-moment $m_{eff}(\psi = \pi/2)$ is always positive for $0 < Q < 1$ and zero only at $Q=1$, a scenario similar to the membrane force $f$. This proves that monodomain nematic strips will always bend toward the incoming light.

The minimum-maximum ratios $f(\pi/2)/f(0)$ and $m_{eff}(\pi/2)/m_{eff}(0)$ in Fig. 5(a), (b) reassure intuitive thoughts originated from (4) regarding impacts of the order parameter $Q$ upon polarization dependence. Both ratios are decreasing functions of $Q$, confirming that a big $Q$ amplifies the polarization dependence as inferred from (4). When $Q=0$ *i.e.* for isotropic samples, the minimums are equal to maximums ---- polarization dependence vanishes; when $Q=1$ *i.e.* for perfectly aligned LCs, the minimums are 0 ---- no photo-contraction nor bend, no photo-response at all at perpendicular illuminations ($\mathbf{e} \perp \mathbf{n}$). When $d/h$ is small, the ratios are close to 1 for most $Q$ ---- polarization dependence is weak ---- and the maximum-minimum gaps widen as $d/h$ rises. This is because in the weak light limit, $n_c$ varies with $\psi$ most abruptly near the surface (refer to $i_0$=0.5 in Fig. 2a), in other words, polarization dependence is predominant for thin strips. For a same given $d/h$, $m_{eff}(\pi/2)/m_{eff}(0)$ is smaller than $f(\pi/2)/f(0)$, that is polarization dependence manifests more on the photo-moment than the membrane force. This is because membrane force shows an average effect over depths, which flattens polarization dependence to some extent, while moment reflects more details about *trans* depletion distribution (Fig. 2a).

The following is a brief sum-up of this subsection for weak light limit. Analytical expressions for membrane force, averaging the photo-contraction over depths, and photo-moment, to describe the variation of the photo-contraction over depths, are obtained. Both prove to be maximum when $\mathbf{e} \parallel \mathbf{n}$ and minimum when $\mathbf{e} \perp \mathbf{n}$ under weak light. Polarization dependence is magnified by order parameter. The analytical expression for the membrane force is used to fit the experimental data in [1].

*3.3 Polarization dependence of photo-moment under strong light illumination*

We have investigated through (3), (6), (7) that $f$ grows with $i_0$~1 monotonically. Under strong illumination, $f$ behaves qualitatively the same as in weak light limit, though $f$ cannot be explicitly expressed as weak light limit does in (9). But the photo-moment (8) shows rather different characteristics as the nonlinearity of photo-moment has been reported for diffused light in [18]. The effective photo-moment is shown against incident light intensity in Fig. 6a, which sees a non-monotonic up and down curve. Recall $n_c$ under different illuminations $i_0$ in Fig. 2a. When light is too weak, $n_c$ or $\varepsilon$ is too small to cause a bending moment or curvature. When light is too strong, *trans* is so highly bleached that a high $\varepsilon$ almost uniform through thickness cannot contribute to bend. The rise for $\psi=0$ in Fig. 6a under weak light is the steepest among all $\psi$ but the demise under strong light is also the fastest. Intersection of $M_{eff}$-$i_0$ curves for different $\psi$ is expected. Such switchover of polarization dependence in $M_{eff}$ can be observed by another view in Fig. 6b, where $M_{eff}$ is plotted against $\psi$ for different $i_0$. At weaker light, $i_0 < i_1^c$, $M_{eff}$ decreases within $\psi \in [0, \pi/2]$, where $\psi$ for $M_{eff}$ to be maximum is $\psi_M = 0$. When $i_0 > i_1^c$, $M_{eff}$ becomes a non-monotonic function of $\psi$, where $\psi_M$ is no longer $0$. When $i_0 > i_2^c$, $M_{eff}$ becomes an increasing function of $\psi$, with $\psi_M = \pi/2$. The two critical light intensity $i_1^c$ and $i_2^c$ are thresholds for the change of monotonicity of $M_{eff}$-$\psi$: monotonically decreasing for $i_0 < i_1^c$, non-monotonic for $i_1^c < i_0 < i_2^c$ and monotonically increasing for $i_0 > i_2^c$. Fig. 6c illustrates $\psi_M$ for maximum curvature, jumping from $\psi_M = 0$ at weak light $i_0 < i_1^c$ to $\psi_M = \pi/2$ at strong light $i_0 > i_2^c$. $i_1^c$ and $i_2^c$ are generally functions of *h/d* though only denoted for *h/d*=1 in Fig. 6b.

Recall $n_c$ in Fig. 2a for an intuitive explanation of the above switchover in $M_{eff}$-$\psi$. Take *h/d* certain value (say *h/d*=1) less than more or less where bleaching fronts locate under a particular strong intensity (say $i_0$=5). When $\psi=0$, $n_c$ drops quickly under weak light; *trans* is most bleached rather uniformly (0<*z/d*<1) above the bleaching front under strong light ---- deep penetration. For other $\psi$, *trans* is highly bleached under strong light as well, but not as highly and evenly as for $\psi=0$. Under strong enough light, while $\psi=0$ hardly gives gradient $n_c$ for curvature, $n_c$ for $\psi = \pi/2$ slightly looks like Beer exponential decay (Fig. 2a), which still gives a significant curvature. That is when $M_{eff}(\psi=\pi/2)$ becomes greater than $M_{eff}(\psi=0)$ and even maximum among all $\psi$. It is also helpful to look at $i_1^c(h/d)$ and $i_2^c(h/d)$, both increasing functions of *h/d* shown in Fig. 6c, d. Under a certain illumination intensity: when sample is thin, light is so deeply penetrated that bleaching fronts are outside of the sample (Fig. 2a), and so *trans* is most bleached at $\psi=0$ as stated above ---- the above switchover happens; when sample is thick, light is as if trapped near the skin and bleaching fronts lie within the sample. If bleaching fronts lie within the sample, the bleaching front for $\psi=0$ is the steepest (Fig. 2a), thereby giving a biggest curvature ---- no switchover happens. In order for the switchover to happen, illumination should be strong enough to penetrate the whole sample, that is why $i_1^c(h/d)$ and $i_2^c(h/d)$ both increase with *h/d*.

A possible analytical demonstration for the switchover of $M_{eff}$-$\psi$ can be done for relatively thin strips, *i.e. h/d*<<1. Using (7) and (8), *f* and $m_{eff}$ can be expanded up to the 1st term in *h/d* polynomials as

$$f = \varepsilon|_{z=0} + \cdots, \tag{13}$$

$$m_{eff} = -\frac{1}{12}\frac{d\varepsilon}{dz}\bigg|_{z=0}\frac{h}{d} + \cdots. \tag{14}$$

$m_{eff}$ is one order higher in *h/d* than *f*. Using (3) and (6), one gets

$$f = \gamma_\varepsilon \frac{i_0\hat{\eta}}{1+i_0\hat{\eta}}, \tag{15}$$

$$m_{eff} = \frac{\gamma_\varepsilon}{12}\frac{i_0\hat{\eta}^2}{(1+i_0\hat{\eta})^3}\frac{h}{d}. \tag{16}$$

Note the 1st term in *f* is just $\varepsilon$, which is proportional to $n_c$ by setting *i*=1 in (3) ---- surface $n_c$. In the thin strip limit, *i.e. h/d*<<1, polarization dependence of *f* is the same as $\varepsilon$ at the surface. Polarization dependence of $m_{eff}$ can be observed by its derivative

$$\frac{dm_{eff}}{d\psi} = -\gamma_\varepsilon \frac{h}{4d}\frac{Qi_0\hat{\eta}\sin 2\psi}{(1+i_0\hat{\eta})^4}(2-i_0\hat{\eta}), \tag{17}$$

where $\psi$ enters $\hat{\eta}$ as well. It is obvious that under weak light $i_0$<<1, the above derivative is negative for $\psi\in[0,\pi/2]$ that is $m_{eff}$ monotonically decreases with $\psi\in[0,\pi/2]$. The $m_{eff}$-$\psi$ monotonicity holds for $i_0 \leq i_1^c := 2/(1+2Q)$ obtained by (17). When $i_1^c < i_0 < i_2^c := 2/(1-Q)$, $m_{eff}$-$\psi$ becomes non-monotonic, with $\psi_M$ given by

$$\cos^2 \psi_M = \frac{1/i_0 - 1/i_2^c}{1/i_1^c - 1/i_2^c}. \tag{18}$$

Note that $i_{1,2}^c$ here are obtained by (17). The above angle $\psi_M$ is also shown in Fig. 6c as a limiting case for very small $h/d$. $i_{1,2}^c$ obtained by (17) agree well with the general curves in Fig. 6d in the $h/d<<1$ limit, where $Q$ is chosen as 0.45. As light increases to $i_0 \geq i_2^c$, $m_{eff}$ becomes monotonically decreasing with $\psi$.

A more direct representation of the effects of photo-moments is photo-deflection (denoted by $\zeta$) of a cantilever. Assuming infinitesimal strain and deflection, the 2$^{nd}$ spatial derivative of photo-deflection is the cantilever curvature, which is proportional to the photo-moment, provided there is no other external mechanical torque:

$$\frac{1}{12} Ewh^3 \zeta_{,xx} = M_{eff}. \tag{19}$$

Then, the photo-deflection spatial distribution function is

$$\zeta = \frac{6 M_{eff}}{Ewh^3} x^2, 0 < x < L. \tag{20}$$

So the bigger the photo-moment, the bigger the photo-deflection in this case. Fig. 7 shows the photo-deflection distribution on a cantilever clamped at $x=0$ shone by light from above. When incident light intensity is relatively weak, the polarization dependence of the photo-moment is trivial (as indicated in Fig. 6b), so $\psi = 0$ gives the biggest deflection as shown in Fig. 7a. As incident light intensity is stronger, the polarization dependence becomes non-monotonic. In Fig. 7b, $\psi = \pi/4$ gives the biggest deflection; in Fig. 7c, $\psi = \pi/2$ gives the biggest deflection while $\psi = 0$ gives the smallest deflection.

A brief conclusion of this subsection is follows. Membrane force under strong light does not show qualitative difference from weak light limit. Photo-moment under strong light shows switchover in polarization dependence, namely parallel illumination does not always trigger maximum bend and perpendicular illumination could give maximum bend. This switchover is explained by the penetration depth of the *cis* fraction/photo-contraction distribution.

4. **Conclusions**

By introducing the polarization dependent photo-isomerization theory proposed by Corbett and Warner [16][23] into modeling thin strips, we have analyzed the polarization dependence of the opto-mechanical behavior of monodomain glassy liquid crystals under polarized light illumination. We have found that:

1) Within a photo-sensitive azo-LC medium, *cis* isomer fraction $n_c(z)$ and photo-contraction $\varepsilon(z)$

   are maximum when **e** ∥ **n** near illumination surface ($z \leq z_1^c$). Below certain depth ($z > z_2^c$), the distributions will be switched over with minimum when $e \parallel n$ and maximum when $\mathbf{e} \perp \mathbf{n}$. *Trans* bleaching fronts lie right below the surface under weak light but beneath a certain depth under strong light.

2) For a strip sample with director along axis, the average contraction or equivalently the membrane force measured by fixing both ends of the sample is found to be always maximum when $e \parallel n$ as observed in [1], fitted by our analytical expression.

3) However, photo-moment $M_{eff}$ related to bending curvature could have a more sophisticated dependence upon the polarization angle due to the strongly nonlinear behavior of the photo-contraction $\varepsilon(z)$. Only at weak light illuminations ($i_0 \leq i_1^c$), $M_{eff}$ has a maximum value when $e \parallel n$ ---- switchover will occur at some stronger light ($i_0 > i_2^c$) with minimum curvature at $e \parallel n$ and maximum at $\mathbf{e} \perp \mathbf{n}$. In between ($i_1^c < i_0 < i_2^c$), the maximum bending moment will be induced by a suitable polarization angle $0 < \psi_M < 90°$. It is also illustrated that the critical intensities for the switchover $i_{1,2}^c$ increase with sample thickness $h$, for switchover requires deep penetration. A thin strip limit is given for analytical demonstration of the counter-intuitive switchover.

Our results indicated that photo-response (contraction and bend) of *trans-cis* glassy LC polymers requires careful modeling and that the polarization dependence is by no means as trivial as expected. Further experiments should be designed to test the above results. In addition, all the results above on strips can be applied to plates or films, in that the effective photo-moment of films in [21] is exactly (8) in this paper. Thus there will be effective photo-moment along the LC orientation corresponding to bend towards light, and always anticlastic bend backward from the light in the transverse direction with an amount of minus Poisson Ratio times (8). Furthermore, given the interest and complexity in photo-response of glassy LC polymers in this paper, more extensive photo-mechanical modeling and response in for instance, LC elastomers, whose directors are mobile, and *trans-cis-trans* isomerization are worth looking forward to.

**Acknowledgements**

The authors acknowledge the support of the National Natural Science Foundation of China (Nos 11461161008，11272092). We thank Dr. John Simeon Biggins for precious suggestions.

**References**

[1] Harvey C L M, Terentjev E M. 2007. Role of polarization and alignment in photoactuation of nematic elastomers. *The European Physical Journal E*, 23(2): 185-189.

[2] Finkelmann H, Nishikawa E, Pereira G G, et al. 2001. A new opto-mechanical effect in solids. *Physical Review Letters*, 87(1): 015501.

[3] Hogan P M, Tajbakhsh A R, Terentjev E M. 2002. UV manipulation of order and macroscopic shape in nematic elastomers. *Physical Review E*, 65(4): 041720.

[4] Warner M, Terentjev E M. 2003. Liquid Crystal Elastomers. *Oxford University Press*.

[5] Ikeda T, Nakano M, Yu Y, et al. 2003. Anisotropic bending and unbending behavior of azobenzene liquid-crystalline gels by light exposure. *Advanced Materials*, 15(3): 201-205.

[6] Yu Y, Nakano M, Ikeda T. 2003. Photomechanics: directed bending of a polymer film by light. *Nature*, 425(6954): 145-145.

[7] Finkelmann H., Kim S.T., Munoz A., Palffy-Muhoray P., and Taheri B. 2001. Tunable mirrorless lasing in cholesteric liquid crystalline elastomers, *Adv. Mater.* 13. 1069–1072.

[8] White T J, Serak S V, Tabiryan N V, et al. 2009. Polarization-controlled, photodriven bending in monodomain liquid crystal elastomer cantilevers. *Journal of Materials Chemistry*, 19(8): 1080-1085.

[9] Lee K M, Smith M L, Koerner H, et al. 2011. Photodriven, flexural–torsional oscillation of glassy azobenzene liquid crystal polymer networks. *Advanced Functional Materials*, 21(15): 2913-2918.

[10] Dunn M L. 2007. Photomechanics of mono-and polydomain liquid crystal elastomer films. *Journal of Applied Physics*, 102(1): 013506.

[11] Dunn M L, Maute K. 2009. Photomechanics of blanket and patterned liquid crystal elastomer films. *Mechanics of Materials*, 41(10): 1083-1089.

[12] Cheng L, Torres Y, Lee K M, et al. 2012. Photomechanical bending mechanics of polydomain azobenzene liquid crystal polymer network films. *Journal of Applied Physics*, 112(1): 013513.

[13] Gaididei Y B, Krekhov A P, Büttner H. 2010. Nonlinear bending of molecular films by polarized light. *Physics Letters A*, 374(21): 2156-2162.

[14] Corbett D, Xuan C, Warner M. 2015. Deep optical penetration dynamics in photobending. *Physical Review E*, 92(1): 013206.

[15] Marrucci L, Paparo D, Abbate G, et al. 1998. Enhanced optical nonlinearity by photoinduced molecular orientation in absorbing liquids. *Physical Review A*, 58(6): 4926.

[16] Corbett D, Warner M. 2006. Nonlinear photoresponse of disordered elastomers. *Physical Review Letters*, 96(23): 237802.

[17] Warner M, Mahadevan L. 2004. Photoinduced deformations of beams, plates, and films. *Physical Review Letters*, 92(13): 134302.

[18] Jin L, Yan Y, Huo Y. 2010. A gradient model of light-induced bending in photochromic liquid crystal elastomer and its nonlinear behaviors. *International Journal of Non-Linear Mechanics*, 45(4): 370-381.

[19] Jin L, Lin Y, Huo Y. 2011. A large deflection light-induced bending model for liquid crystal elastomers under uniform or non-uniform illumination. *International Journal of Solids and Structures*, 48(22): 3232-3242.

[20] You Y, Xu C, Wang B, et al. 2011. Photo-actuated bending of chromatic liquid crystal polymer strips and laminates. *International Journal of Smart and Nano Materials*, 2(4): 245-260.

[21] You Y, Xu C, Ding S, et al. 2012. Coupled effects of director orientations and boundary conditions on light induced bending of monodomain nematic liquid crystalline polymer plates. *Smart Materials and Structures*, 21(12): 125012.

[22] Lin Y, Jin L, Huo Y. 2012. Quasi-soft opto-mechanical behavior of photochromic liquid crystal elastomer: Linearized stress–strain relations and finite element simulations. *International Journal of Solids and Structures*, 49(18): 2668-2680.

[23] Corbett D, Warner M. 2008. Polarization dependence of optically driven polydomain elastomer mechanics. *Physical Review E*, 78(6): 061701.

[24] Corbett D, Warner M. 2009. Changing liquid crystal elastomer ordering with light–a route to opto-mechanically responsive materials. *Liquid Crystals*, 36(10-11): 1263-1280.

**Figures**

Fig. 1

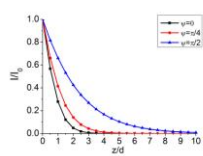

Fig. 2

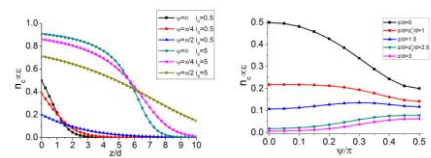

Fig. 3

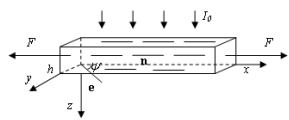

Fig. 4

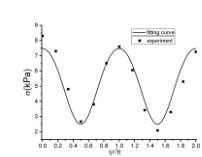

Fig. 5

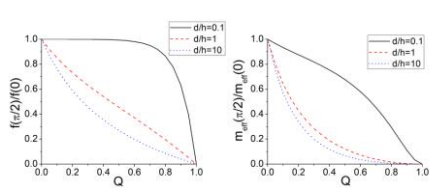

Fig. 6

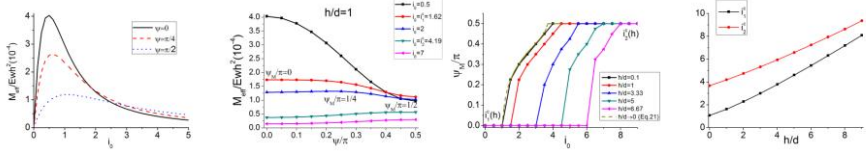

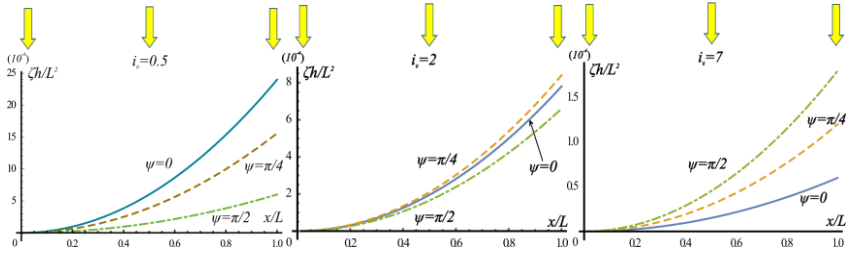

Fig. 7

Fig. 1. Light intensity $I/I_0$ as a function of propagation depth $z/d$ for $\psi=0$, $\psi=\pi/4$ and $\psi=\pi/2$.

Fig. 2.

(a) *Cis* fraction $n_c$ (proportional to photo-contraction $\varepsilon$) as a function of propagation depth $z/d$ for $\psi=0$, $\psi=\pi/4$ and $\psi=\pi/2$. Two dimensionless light intensities $i_0=0.5$ and $i_0=5$ are given.

(b) *Cis* fraction $n_c$ (proportional to photo-contraction $\varepsilon$) as a function of polarization angle $\psi/\pi$ for different illumination depths $z/d$.

Parameter definitions: $i_0=\tau_{ct}\eta_0 I_0$, $d=(\gamma_t\eta_0)^{-1}$.

Fig. 3. Illustration of a monodomain LC glassy strip under illumination.

Fig. 4. Engineering photo-stress against polarization angle $\psi/\pi$. Line: fitting curve (9); dots: experiment in [1].

Fig. 5.

(a) Membrane force minimum-maximum ratios $f(\pi/2)/f(0)$ as a function of order parameter $Q$ for different strip thickness $h/d$.

(b) Photo-moment minimum-maximum ratios $m_{eff}(\pi/2)/m_{eff}(0)$ as a function of order parameter $Q$ for different strip thickness $h/d$.

Fig. 6.

(a) Photo-moment as a function of illumination intensity $i_0$ for $\psi=0$, $\psi=\pi/4$ and

$\psi = \pi/2$.

(b) Photo-moment as a function of polarization angle $\psi/\pi$ for different illumination intensity $i_0$ when $h/d = 1$.

(c) Optimal polarization angle $\psi_M$ for maximum photo-moment as a function of illumination intensity $i_0$ for different strip thickness $h/d$.

(d) Critical illumination intensities for the switchover in photo-moment.

Fig. 7. Cantilever deflection under different polarization angles ($h/d$=1)
- (a) Light intensity $i_0$=0.5.
- (b) Light intensity $i_0$=2.
- (c) Light intensity $i_0$=7.